\documentclass[12pt]{article}
\usepackage[body={18cm, 23cm, centered}]{geometry}
\usepackage{amsmath,amssymb,amsthm,amscd,cite,comment,amsfonts,indentfirst,color,setspace,bbold,MnSymbol,arydshln}
\usepackage{mathtext,marvosym,textcomp}
\usepackage[makeroom]{cancel}
\usepackage{mathtools}

\usepackage[debug,pageanchor=false]{hyperref}
\hypersetup{colorlinks=true,linktocpage,breaklinks,
	urlcolor=blue,
	linkcolor=red,
	citecolor=blue
}

\numberwithin{equation}{section}

\selectfont

\def\a{\alpha} 
\def\b{\beta} 
\def\g{\gamma} 
\def\d{\delta} 
\def\e{\epsilon}

\def\h{\eta}

\def\l{\lambda} 
\def\m{\mu}
\def\n{\nu} 
 
\def\r{\rho}

\def\s{\sigma} 
\def\t{\tau}  
\def\f{\phi}
 
\def\w{\omega}

\def\G{\Gamma}

\def\I{\iota} 
 
\def\L{\Lambda}

\def\S{\Sigma}

\def\ra{\rangle}

\def\fr{\frac}  \def\dt{\partial}

\def\mc{\mathcal}

\def\mC{\mathcal{C}}

\def\mF{\mathcal{F}}
\def\mG{\mathcal{G}}
\def\mH{\mathcal{H}}

\def\mH{\mathcal{H}}

\def\tc{\tilde{c}}

\def\tk{\tilde{k}}
\def\tx{\tilde{x}}

\def\tK{\tilde{K}}

\def\tY{\tilde{Y}}
\def\tmG{\tilde{\mG}}

\def\XX{\mathbb{X}}
\def\RR{\mathbb{R}}

\def\hdt{\hat{\dt}}

\def\tmG{\tilde{\mG}}

\def\cB{\mathcal{B}}
\def\cC{\mathcal{C}}
\def\cG{\mathcal{G}}

\def\rO{\mathrm{O}}

\begin{document}
\renewcommand{\refname}{\begin{center}References\end{center}}
	
\begin{titlepage}
		
	\vfill
	\begin{flushright}

	\end{flushright}
		
	\vfill
	
	\begin{center}
		\baselineskip=16pt
		{\Large \bf 
		The invariant action for solitonic 5-branes
		}
		\vskip 1cm
			Edvard T. Musaev$^{a,b}$\footnote{\tt musaev.et@phystech.edu}, Jeffrey P. Molina$^{a}$\footnote{\tt paredes.md@phystech.edu},
		\vskip .3cm
		\begin{small}
			{\it 
				$^a$Moscow Institute of Physics and Technology,
			    Institutskii per. 9, Dolgoprudny, 141700, Russia,\\
				$^b$Kazan Federal University, Institute of Physics, Kremlevskaya 16a, Kazan, 420111, Russia\\
			}
		\end{small}
	\end{center}
		
	\vfill 
	\begin{center} 
		\textbf{Abstract}
	\end{center} 
	\begin{quote}
            We construct the full effective action including DBI and WZ terms for solitonic 5-branes covariant under T-duality. The result is a  completion of results known in the literature  to a full T-duality covariant expression. The covariant WZ action includes previously omitted R-R terms. The obtained full covariant effective action reproduces the one obtained by S-duality from the D5-brane upon the correct choice of the covariant charge.
	\end{quote} 
	\vfill
	\setcounter{footnote}{0}
\end{titlepage}
	
\clearpage
\setcounter{page}{2}

\tableofcontents

\section{Introduction}
Branes, {\it i.e.}  non-perturbative extended objects in string theory, have played,  since the discovery of string dualities, an ubiquitous role in the field. D-branes, defined as the endpoints of open strings, are crucial in the construction of orientifold models \cite{Sagnotti:1987tw,Pradisi:1988xd,Bianchi:1990yu,Polchinski:1995mt} (for a review see \cite{Angelantonj:2002ct}), while in F-theory models \cite{Vafa:1996xn}, in which the IIB string coupling varies along the internal manifold, one is forced to introduce additional 7-branes related to the D7-branes by S-duality \cite{Sen:1996vd}. Finally, branes are at the heart of the original formulation AdS/CFT correspondence \cite{Maldacena:1997re} and all its generalizations, and in this context they have been essential in our ever-improving understanding of superconformal field theories in various dimensions. 

What becomes clear though putting all these ingredients together is that branes are not enough. As an example \cite{Hellerman:2002ax}, one can relate by duality an F-theory model to a model in IIA where there is a non-trivial monodromy for the complex modulus describing the internal volume and the internal $B$-field. This corresponds to the presence of both NS5-branes and branes obtained by the latter performing two T-dualities in the transverse directions. Such branes, named $5_2^2$-branes in the literature \cite{Obers:1998fb},\footnote{The lower number denotes how the tension scales with respect to the inverse sting coupling, while the upper number denotes the number of isometries.} are intrinsically non-geometric \cite{deBoer:2012ma}, in the sense that the corresponding supergravity solution can not be uplifted to ten dimensions. 

The fact that string dualities imply in general the presence of {\it exotic branes}, that is branes  in lower dimensional theories that have no origin in ten dimension, had already been observed earlier \cite{Giveon:1998sr,Elitzur:1994ri,Obers:1998fb}, and corresponding solutions had been constructed in \cite{Lozano-Tellechea:2000mfy}. In \cite{Eyras:1998hn} the effective action of the IIA KK monopole 
was constructed by T-dualising the IIB NS5-brane effective action, that in turn is obtained by S-duality from the D5. This was later generalized in \cite{Chatzistavrakidis:2013jqa,Kimura:2014upa} to obtain the effective actions of additional exotic branes, and in particular the IIB $5_2^2$-brane. In a series of papers \cite{Bergshoeff:2010xc, Bergshoeff:2011qk, Bergshoeff:2011zk, Bergshoeff:2013sxa, Bergshoeff:2019sfy}, the Wess-Zumino (WZ) term of the effective actions of branes in lower dimensions was derived using the transformation properties of the potentials under T-duality and gauge symmetry. In order to make T-duality manifest, the world-volume potentials are treated in a democratic formulation, in which each potential generally appears together with its magnetic dual. It is worth mentioning that this method does not fix terms that are gauge invariant and do not mix with the other terms by T-duality. We will come back to this point later in the paper.

A natural question that one can ask is how the effective actions that one gets from S-duality starting from the D5-brane are related to the effective actions in \cite{Bergshoeff:2019sfy}, which are manifestly covariant under T-duality. We will answer this question in the first part of the paper, providing the full effective action for the IIB NS5-brane in a formulation which is fully democratic for both the world-volume and target space potentials. The result is then applied to obtain the full effective action for the IIA KK-monopole using T-duality.

The natural framework in which T-duality is manifestly realized is Double Field Theory (DFT) \cite{Siegel:1993th,Siegel:1993xq,Hohm:2010pp} (for a comprehensive review see \cite{Aldazabal:2013sca,Berman:2013eva}). This is a theory whose fields belong to irreducible representations of the O$(10,10;\RR)$ group, that includes global coordinate transformations, constant shifts of the Kalb--Ramond field and T-duality transformations. The field content includes the generalized metric $\mH_{MN}$ parametrizing the coset space O$(10,10)/\rO(1,9)\times \rO(9,1)$ and the invariant dilaton $d$, which formally depend on a doubled set of coordinates $\XX^M = \{x^m, \tx_m\}$ with $M=1,\dots,20$ and $m=0,\dots,9$. Dynamics of the theory is given by an action invariant under generalized Lie derivative related to transformations of coordinates on the doubled space. Consistency of the algebra of generalized diffeomorphisms requires to impose additional condition on the space of functions, that can be schematically written as 
\begin{equation}
    \label{eq:sc}
    \h^{MN}\dt_M \otimes \dt_N = 0,
\end{equation}
where $\h^{MN}$ is the invariant tensor of the orthogonal group. This reduces the space of allowed functions to only those that depend on  half of the coordinates. Conventionally one considers functions that depend on either $x^i$ or $\tx_i$ for each pair of coordinates for a given $i$. Although more general  dependence is possible, we do not consider it here (see \cite{Astrakhantsev:2021rhj} for examples obtained by non-abelian fermionic T-duality). There are however subtleties related on whether a given $x^i$ is understood as a geometric or non-geometric coordinate, which we discuss in more details further in the text.

There exists a choice of solution to the section constraint \eqref{eq:sc} that reduces DFT to the standard 10-dimensional supergravity. Equations for the latter allow extremal black brane solutions whose sources can be identified with branes of string theory. On the other hand all supergravity solutions that belong to a given T-duality orbit uplift to a single solution of DFT equations, e.g. the orbit $5_2^0-\dots -5_2^4$ uplifts to the so-called DFT-monopole \cite{Berman:2014jsa,Berkeley:2014nza,Bakhmatov:2016kfn} (for review of description of exotic branes in terms of extended field theories see \cite{Musaev:2019zcr,Berman:2020tqn}). Naturally one is interested in uplifting the correspondence source--solution to a T-duality covariant framework and consider a full action 
\begin{equation}
    \label{eq:fullschem}
    S_{\mathrm{full}} = S_{\mathrm{brane}}[\dt_a Y^M, k_\alpha^M] + S_{\mathrm{DFT}}[\mH_{MN},d,\chi] \ .
\end{equation}
Here $S_{\mathrm{DFT}}[\mH_{MN},d,\chi]$ is the action for double field theory (where  $\chi$ is a DFT spinor describing the R-R fields \cite{Hohm:2011zr,Hohm:2011dv}) and $S_{\mathrm{brane}}[\dt_a Y^M,k_\alpha]$ is an action for a covariant $n$-dimensional object. Its embedding in the  doubled space is defined by functions $Y^M= Y^M(\s^a)$ and a set of  Killing vectors $k_\alpha^M$ encoding information on isometric directions. The latter are necessary due to the section condition, that requires independence on at least a half of the coordinates. Finally, $\s^a$ with $a=0,\dots,n$ parametrise the world-volume of the brane.

To end up with a single object embedded in the extended space sourcing the single DFT solution it is natural to require for the first term in \eqref{eq:fullschem} the same symmetries that the second term possesses. Several steps in this direction  have been made for T-duality brane orbits with given dependence of tension on $g_s$, which is a natural approach since T-duality does not mix them. To be mentioned are the works \cite{Albertsson:2008gq,Albertsson:2011ux} considering covariant description of D$p$-branes in terms of maximally isotropic submanifolds in the doubled space, the work \cite{Marotta:2022tfe} describing D-branes in terms of Born sigma models (see \cite{Sakatani:2020umt} for a similar approach to branes of M-theory), the works \cite{Kimura:2015qze,Blair:2017hhy} focusing on $g_s^{-2}$ (solitonic) branes and \cite{Bergshoeff:2019sfy} addressing double field theory description of Wess-Zumino terms for various branes with tension down to $g_s^{-4}$. Relevant for the present discussion are the works \cite{Blair:2017hhy} and \cite{Bergshoeff:2019sfy}. In the former the kinetic term and the leading WZ terms of the IIB NS5-brane action in a DFT formulation was constructed based on the notion of generalized Killing vector associated to the DFT-monopole. In \cite{Bergshoeff:2019sfy} Wess-Zumino terms for $g_s^{-2}$-branes in a DFT formulation have been presented, which however i) are not in the form that reproduces the standard WZ action obtained by S-duality of the D5-brane; ii) do not complete the DBI action of \cite{Blair:2017hhy} since a different approach has been used. The second issue can be relatively easy cured by rewriting the result using the notion of generalized Killing vectors for a DFT-monopole. The first issue is more important since the WZ terms have been constructed based on the expression of \cite{Bergshoeff:2011zk} that missed R-R terms which are T-duality invariant by themselves and hence are not seen by the analysis.

 In this paper we present the full covariant effective action for the DFT-monopole using  the approach of  \cite{Blair:2017hhy}, which reproduces exactly the known expressions for the NS5-brane and its T-duality partners (upon smearing). In Section \ref{sec:nots} we set the notations for the fields entering the effective actions of the D5 and NS5 branes. We rewrite the equations for the DBI field $c_1$ following from the NS5-brane action to introduce its world-volume magnetic dual $c_3$, and present the complete Wess-Zumino action in notations that facilitate further rewriting in the DFT language. Section \ref{sec:cov} starts with a short reminder of the double field theory formalism followed by an introduction of the fully covariant action \eqref{eq:S_full} and  a discussion of the subtleties related to the notion of a non-geometric coordinate. We show that for certain choices of generalized isometries one successfully reproduces exactly the full effective action for the NS5-brane including all R-R terms missed in previous analyses. We show that a change in the set of generalized Killing vectors which corresponds to a T-duality transformation precisely gives the covariant Buscher rules. We illustrate this using  the effective action of KK5A. Finally in Section \ref{sec:disc} we recompile the results of the paper, elaborate on the importance of generalized isometric directions and discuss relations between the results presented here and  those in the literature.

\section{Effective action for Type II 5-branes}

\label{sec:nots}

\subsection{Conventions for NSNS and R-R potentials}

Let us start with reviewing different notations for supergravity and brane world-volume fields being used in the literature and setting up the notations we stick to further in the text. Important for further discussion is to distinguish between fields that transform covariantly under T-duality and those that do not. The R-R fields that are commonly used in the supergravity literature are the $p$-form potentials $C_p$ that transform as
\begin{equation}
\delta C = e^{-B_2} d \lambda \ , 
\end{equation}
where $C$ is the formal sum of the forms $C_p$. The resulting gauge invariant field strength is
\begin{equation}
    G_{p+1} = d C_p + H_3 \wedge C_{p-2} \ .
\end{equation}
In this basis the D-brane WZ term is simply
\begin{equation}
S_{WZ}^{{\text D}p} = T_p\int_{\S_{p+1}} e^{\mathcal{F}_2} C
\end{equation}
which is the standard notation in supergravity. Here $\mathcal{F}_2 = d A_1 + B_2$ is the field strength of the world-volume Born--Infeld fields with gauge transformations defined by $\delta A_1 = -\L_1$ and $\delta B_2 = d \L_1$, and $\S_{p+1}$ denotes the $p+1$-dimensional world-volume. In this basis the R-R potentials do not transform under $\L_1$, and T-duality transformations  mix $C_p$ with $B_2$.

On the contrary, in the T-duality basis R-R potentials   $\cC= e^{B_2} C$ transform covariantly under T-duality transformations simply as components of an $\rO(10,10)$ spinor. Their gauge transformations are given by
\begin{equation}
    \delta \cC=  d \lambda + d \L_1 \wedge \cC  \ ,
\end{equation}
and the gauge invariant WZ term for the D$p$--brane reads [probably we should be careful with $p$'s]
\begin{equation}
    S_{WZ}^{{\text D}p} = T_p\int_{\S_{p+1}} e^{F_2} \cC
\end{equation}
where $F_2 = dA_1$. 

We can work out what happens under S-duality. The potential $C_2$ transforms like $B_2$, which implies that $C_2$ and $B_2$ form a doublet
\begin{equation}
    \begin{pmatrix}
    B_2 \\ C_2
    \end{pmatrix}
\end{equation}
and under S-duality transformation
\begin{equation}
    S =\begin{pmatrix}0 & -1 \\ 1 & 0 \end{pmatrix}
\end{equation}
one gets 
\begin{equation}
    B_2 \rightarrow -C_2 \qquad C_2 \rightarrow B_2 \ . 
\end{equation}
Both the  potentials $\cC_4$ and $C_4$ are not invariant under S-duality, as can be deduced from their gauge transformations. However, defining
\begin{equation}
    \begin{aligned}
        \mathcal{A}_4 &= C_4 + \frac{1}{2} C_2 \wedge B_2 \\
        &= \cC_4 -\frac{1}{2} \cC_2 \wedge B_2
    \end{aligned}
\end{equation}
one gets gauge transformation
\begin{equation}
    \delta \mathcal{A}_4 = d \lambda_3 + \frac{1}{2} \big(C_2 \wedge d \L_1 - B_2 \wedge d \lambda_1 \big) 
\end{equation}
which is invariant under S-duality.  This implies the following transformation of the 4-form R-R potential
\begin{equation}
    C_4 \rightarrow C_4 + C_2 \wedge B_2 \ .
\end{equation}
The D5-brane, which is S-dual of the NS5B-brane, interacts with the 6-form R-R-potential. To define the 6-form NSNS potential $\cB_6$ we first consider the potential
\begin{equation}
    \begin{aligned}
        \mathcal{A}_6 &= C_{6} -\frac{1}{6} C_2 \wedge B_2  \wedge B_2 \\
        &= \cC_6 -\cC_4 \wedge B_2 + \frac{1}{3} \cC_2 \wedge B_2 \wedge B_2         
    \end{aligned}
\end{equation}
whose gauge transformation
\begin{equation}
    \delta \mathcal{A}_6 = d \lambda_5 - B_2 \wedge d \lambda_3 -\frac{1}{3}B_2 \wedge \big(C_2 \wedge d \L_1 - B_2 \wedge d \l_1 \big) 
\end{equation}
is manifestly covariant under S-duality. Finally, consider the S-duality doublet which includes both the R-R and NSNS 6-forms
\begin{equation}
    \begin{pmatrix}
    \mathcal{A}_6 \\ \mathcal{B}_6
    \end{pmatrix}
\end{equation}
and read off the following gauge transformations
\begin{equation}
    \delta \mathcal{B}_6 = d \L_5 -d \lambda_3 \wedge C_2 
    -\frac{1}{3}C_2 \wedge (C_2 \wedge d \L_1 - B_2 \wedge d \lambda_1 ) \ .
\end{equation}
On the other hand NS5B-brane is the magnetic dual of the fundamental string, which implies that the potential $\cB_6$ should be 
 compared with the potential $B_6$ that one gets from the electromagnetic duality relation with $B_2$, using the field equations for the Type IIB supergravity action 
\begin{equation}
    \label{eq:SIIB}
    \begin{aligned}
        S_{IIB}  = \kappa \int& e^{-2\phi}\Big(\star R +\fr14 d\phi \wedge \star d\phi+ \fr12 H_3\wedge  H_3\Big), \\
        &+\fr12 G_1 \wedge \star  G_1+\fr12 G_3 \wedge \star  G_3+\fr14 G_5 \wedge \star  G_5 - \fr12 dC_2 \wedge dC_4 \wedge B_2,
    \end{aligned}
\end{equation}
where $G_5=\star G_5$. Equations of motion for the field $B_2$ can be written as Bianchi identities for $H_7 = e^{-2\phi}\star H_3$, that gives
\begin{equation}
    H_7 = dB_6 - C_0 \wedge \star G_3 - \fr12 C_2 \wedge C_2 \wedge H_3 +  dC_2 \wedge C_4.
\end{equation}
Such defined 7-form field strength is invariant under gauge transformations of the 6-form magnetic potential
\begin{equation}
    \delta B_6 = d \L_5 -d \lambda_3 \wedge C_2 + d \lambda_1 \wedge B_2 \wedge C_2 \ .
\end{equation}
This potential has been used in the works \cite{Kimura:2014upa,Chatzistavrakidis:2013jqa} to define Wess-Zumino actions for NS 5-branes. Comparing gauge transformations one finds the relation
\begin{equation}
    \mathcal{B}_6 = B_6 -\frac{1}{3} B_2 \wedge C_2 \wedge C_2 \ ,
\end{equation}
which is consistent with the  the S-duality map
\begin{equation}
    C_6 \rightarrow - B_6 + \frac{1}{2} B_2 \wedge C_2 \wedge C_2. 
\end{equation}

Note however, that neither $B_6$ nor $\cB_6$ are  convenient magnetic potentials to write the Wess-Zumino term for the 5-brane. The reason is simply that their gauge transformations contain the electric gauge potential $B_2$ whose relation to $B_6$ is highly non-trivial. To override this one defines the potential
\begin{equation}
    D_6 = B_6 +\frac{1}{2} C_6 \wedge C_0 + \frac{1}{2} C_2 \wedge C_4
\end{equation}
whose gauge transformations can be written completely in terms of the T-duality covariant R-R potentials $\cC$:
\begin{equation}
    \delta D_6 = d \L_5 +\frac{1}{2} [ d \lambda_5 \cC_0 - d \lambda_3 \cC_2 + d \lambda_1 \cC_4 ]  \ .
\end{equation}
This proves that the potential $D_6$ transforms covariantly under T-duality. Indeed, using the results of \cite{Bergshoeff:2019sfy} and  remembering  that $\cC$ is covariant under T-duality, we can write this relation as
\begin{equation}
    \delta D^{MNPQ} = \partial_R \xi^{RMNPQ} + \bar{\cC} \Gamma^{MNPQ} \Gamma^R\partial_R \lambda
\end{equation}
where $\cC$ and $\lambda$ are chiral spinors of $\rO(10,10)$. In terms of the covariant magnetic potential $D_6$ the 7-form field strength can be written as
\begin{equation}
    H_7 = dD_6 + \fr12 C_0 \wedge G_7-\fr12 C_2 \wedge G_5+\fr12 C_4 \wedge G_3-\fr12 C_6 \wedge G_1,
\end{equation}
where $G_7 = -\star{}G_3$, that follows from equations of motion for the field $C_2$.

We can use all the field redefinitions above to map the D5 brane action to the NS5B brane action written in a form in which we can apply T-duality. For that we define invariant world-volume field strengths for potentials $c_p$ 
\begin{equation}
    \label{eq:gaugeWV}
    \begin{aligned}
        \cG &  = e^{-B_2}d\tc + C= e^{-B_2}(d\tc + \cC) = e^{-B_2}\tmG,\\
        \d \tc_p & = -\l_{p+1} + d\L_1 \wedge \tc_{p-2},
    \end{aligned}
\end{equation}
where we use the same notations for the T-duality covariant potentials $\tc_p$ and field strengths $\tmG_p$ as in \cite{Blair:2017hhy}. However, it appears that the DFT construction of the effective action prefers world-volume potentials defined in a different way such that
\begin{equation}
    \mG_p = dc_p + H_3 \wedge c_{p-3} +C_p,
\end{equation}
where $c_p = e^{-B_2} \tc_p$ are world-volume potentials that are not covariant under T-duality. Such probably unexpected result of preferring non-covariant expressions by a covariant formalism is related to the non-covariance of the factor $e^{B_2}$ in the D5-brane action. This problem has been already faced in the paper \cite{Bergshoeff:2019sfy} when constructing covariant actions for D-branes. Here using such prefactor is avoided by making use of projected world-volume derivatives $\hat{d}$, which contain pull-back of $B_{2}$ inside (see all definitions below). The section condition then requires to have additional derivative $d$ acting on $\hat{d}$ eventually leading to $H_3$ in all expressions.

Given the gauge transformations above, the  gauge invariant Wess-Zumino term for the NS5B brane must be of the form \cite{Bergshoeff:2011zk}
\begin{equation}
    \label{eq:WZgauge0}
        \begin{aligned}
            S_{WZ}^{NS5B} &\sim \int D_6 - \fr12\cG_0 \wedge C_6 + \fr12\cG_2 \wedge C_4 - \fr12\cG_4 \wedge C_2 + \ldots
        \end{aligned}
\end{equation}
where ellipses denote possible additional terms, that are gauge invariant on their own. It is possible to determine the form of at least a subset of these by making use of T-duality covariance arguments. Indeed, it is straightforward to see, that for the  terms of the form $\cG\wedge C$ in \eqref{eq:WZgauge0} to transform covariantly under T-duality the term $\fr12\cG_6\wedge C_0$ must be added. Then the above can be rewritten in the form
\begin{equation}
    \label{eq:WZgauge1}
        \begin{aligned}
            S_{WZ}^{NS5B} &\sim \int D_6 - \fr12\tmG_0 \wedge \cC_6 + \fr12\tmG_2 \wedge \cC_4 - \fr12\tmG_4 \wedge \cC_2 +\fr12\tmG_6 \wedge \cC_0+ \ldots,
        \end{aligned}
\end{equation}
which is completely covariant. Below we show, that there are more gauge invariant terms in the Wess-Zumino action, some of which are of higher power in R-R fields.

\subsection{Effective action for the NS5B-brane}

For completeness of the narration it is convenient to start with the effective action for the D5-brane and perform S-duality transformation to obtain that for the NS5-brane. Hence, we write 
\begin{equation}
    \begin{aligned}
        S^{D5} =& -T_5 \int d^6 \s e^{-\f}\sqrt{-\det (G_{ij} + \mF_{ij})} \\
            &+T_5  \int_{M_6} C_6 + C_4 \wedge \mF_2 + \fr12 C_2 \wedge \mF_2 \wedge \mF_2 + \fr1{3!} C_0 \mF_2 \wedge \mF_2 \wedge \mF_2 .
    \end{aligned}
\end{equation}
where  $G_{ij}$ and $B_{ij}$ fields are the pull-backs of the space-time fields. The indices run along the six directions of the worldvolume and we set $2\pi\alpha' = 1$ since this dimensionful factor can be reinserted by dimensional analysis. We summarize S-duality rules discussed in the previous section as follows (see also  \cite{Kimura:2014upa} for the same notations)
\begin{equation}
    \begin{aligned}
        &\tau \to-\fr{1}{\tau}, && B_2 \to -C_2, \\
        &C_2\to B_2, && G \to |\tau|\, G,\\
        &C_4 \to C_4 + C_2\wedge B_2, && C_0 \to |\t|^{-2}C_0,\\
        & C_6 \to -B_6 +\fr12 B_2\wedge C_2\wedge C_2, && A_1 \to -\tc_1,
    \end{aligned}
\end{equation}
where $\tau = C_0 + ie^{-\phi}$ is the complex axio-dilaton field. Then the effective action of the type IIB NS5-brane is\cite{Eyras:1998hn}
\begin{equation}
     \label{eq:WZ_NS5}
    \begin{aligned}
        S^{NS5B} = &-T_{5}\int d^6\xi e^{-2\phi}\sqrt{1+e^{2\phi}C_0^2}\sqrt{-\det\left( G_{ij} - \fr{e^{\phi}}{\sqrt{1+e^{2\phi}C_0^2}} \mathcal{G}_{ij} \right)}\\
    &-T_5  \int_{M_6} B_6 -\fr12 B\wedge C_2\wedge C_2 + (C_4 + C_2 \wedge B_2 ) \wedge \mG_2 - \fr12 B_2 \wedge \mG_2 \wedge \mG_2 + \fr1{3!} |\t|^{-2} C_0 \mG_2 \wedge \mG_2 \wedge \mG_2 .
    \end{aligned}
\end{equation}
The tension of the NS5-brane is defined by $T_{NS5} = g^{-1}T_{D5}$.

To write the Wess-Zumino term above in the formulation democratic w.r.t. the world-volume potentials $\tc_p$, we define the potential $\tc_3$ electromagnetically dual to $\tc_1$. Equations of motion for $\tc_1$ following from the NS5-brane action in its current form are pretty involved due to the square root in the action. To override this we use the same trick as in \cite{AbouZeid:1998he} and rewrite the DBI part of the action in the form quadratic in $\cG_{ij}$. Hence, we have 
\begin{equation}
    S_{DBI}^{NS5} = -T_5 \int d^6\s |\t|e^{-\f}\sqrt{-\det (G_{ij} - |\t|^{-2} \mG_{ij})}= - T_5 \int d^6\s |\t|e^{-\f} (-G)^{\fr14}(-\hat{G})^{\fr14}, \label{DBINS5}
\end{equation}
where $\hat{G} = \det \hat{G}_{ij}$ and we define
\begin{equation}
    \hat{G}_{ij} = G_{ij} -|\t|^{-4} \mG_{i}{}^k\mG_{kj}.
\end{equation}
To rewrite this action in the Howe--Tucker form we introduce a world-volume metric $\g_{ij}$
 \begin{equation}
     S^{NS5}_{DBI} = - \fr14\left( \fr{2}{\a}\right)^{-\fr{1}{2}}T_5 \int d^6\s e^{-\f}(-\g)^{\fr14}(-G)^{\fr14}\left(\g^{ij}\hat{G}_{ij} - \a\right),
 \end{equation}
where $\a=$ const is a cosmological constant, $\g = \det \g_{ij}$ and $\g^{ij}\g_{jk}=\d^i{}_k$. Equations of motion for the world-volume metric then imply
\begin{equation}
    \begin{aligned}
         \g_{ ij } & = \fr2\a\hat{G}_{ij}\equiv\fr{2}{\a}\big(G_{ij} - |\t|^{-4}\mG_i{}^k \mG_{kj}\big),\\
          \a & = \fr13\g^{ij}\hat{G}_{ij}.
    \end{aligned}
\end{equation}
Since $\g_{ij}$ is proportional to $\hat{G}_{ij}$ the contraction $\g^{ij}\hat{G}_{ij}$ is indeed constant on equations of motion of the world-volume metric.

Now we vary the action $S_{DBI}^{NS5}$ w.r.t. the world-volume 1-form field $\tc_i$ to obtain contribution $EoM_{DBI}$ to the full equations of motion, coming from the DBI part of the action:
\begin{equation}
    EoM_{DBI} = \sqrt{\fr{\a}{2}} \sqrt{-\g} T_5 \nabla_i \Big( |\t|^{-4} e^{-\f} \tilde{\mG}^{i j} \Big).
\end{equation}
Here the covariant derivative is defined to be consistent with the world-volume metric $\g_{ij}$, and the components of the 2-form $\tilde{\mG}= \fr12 \tilde{\mG}_{ij}d\s^i\wedge d\s^j$ are defined as
\begin{equation}
    \tilde{\mG}^{ij} =(-\g)^{-\fr14}(-G)^{\fr14}\g^{k[i}\mG_k{}^{j]}.
\end{equation}
Note that $\mG_i{}^j \equiv \mG_{ik}G^{jk}$. Defining a 4-form $\tilde{\mF}_4$ with components
\begin{equation}
    \tilde{\mF}_{ijkl} = \sqrt{\fr{ \a}{8}}\sqrt{-\g}\e_{ijklmn}\tilde{\mG}^{mn},
    \label{eq:4_form}
\end{equation}
we can write contribution $EoM_{DBI}$ to the equations of motion in the following suggestive form
\begin{equation}
    EoM_{DBI} =  T_5 \star_6  d \Big( |\t|^{-4}e^{-\f} \tilde{\mF}_4 \Big),
\end{equation}
where Hodge duality is taken w.r.t. the world-volume metric $\g_{ij}$. This together with the contribution from the Wess-Zumino term gives the following equations for the $\tc_1$ potential
\begin{equation}
    d\left(|\t|^{-4}e^{-\f} \tilde{\mF}_4 - C_4+B_2\wedge d\tc_1 + \fr12 |\t|^{-2} C_0 \mG_2\wedge \mG_2\right)=0.
\end{equation}
At least locally this can be solved as
\begin{equation}
    |\t|^{-4}e^{-\f} \tilde{\mF}_4 + \fr12 |\t|^{-2} C_0 \mG_2\wedge \mG_2= d\tc_3-B_2 \wedge d \tc_1 +C_4 
\end{equation}
for world-volume potentials $\tc_1$ and $\tc_3$ that transform covariantly under T-duality. Hence, we obtain the duality relation defining the world-volume magnetic potential for $\tc_1$
\begin{equation}
    \mG_4 = |\t|^{-4}e^{-\f} \tilde{\mF}_4+ \fr12 |\t|^{-2} C_0 \mG_2\wedge \mG_2.
\end{equation}

Let us now rewrite the Wess-Zumino term \eqref{eq:WZ_NS5} for the NS5B-brane obtained by S-duality of that for the D5-brane in the form of \eqref{eq:WZgauge0} that is explicitly gauge invariant \begin{equation}
    \begin{aligned}
        &S_{WZ}^{NS5}=\\
        & = -T_5 \int_{M_6} D_6 + \fr 12 C_2 \wedge C_4 - \fr12 C_6 C_0 + C_4 \wedge dc_1 - \fr12 B \wedge dc_1 \wedge dc_1    +  \fr1{3!} |\t|^{-2} C_0 \mG_2 \wedge \mG_2 \wedge \mG_2 \\
        &= - T_5 \int_{M_6}D_6 + \fr12\mG_2 \wedge \mG_4 +\fr12C_4 \wedge dc_1 - \fr12C_2 \wedge  dc_3 -\fr12 C_2 \wedge H_3 \wedge c_1 - \fr12 C_6 C_0 + \fr1{3!} |\t|^{-2} C_0 \mG_2 \wedge \mG_2 \wedge \mG_2 \\
        &=- T_5 \int_{M_6}D_6 + \fr12 \mG_2 \wedge \mG_4-  \fr12C_6 C_0 +\fr12\mG_2 \wedge C_4 -\fr12 \mG_4 \wedge  C_2 + \fr1{3!} |\t|^{-2} C_0 \mG_2 \wedge \mG_2 \wedge \mG_2.
    \end{aligned}
\end{equation}
Finally, adding and subtracting $\mG_6 \wedge C_0$ we have
\begin{equation}
    \begin{aligned}
        S_{WZ}^{NS5}   =- T_5 \int_{M_6}&D_6- \fr12  \mG_0 \wedge C_6  +\fr12 \mG_2 \wedge C_4 - \fr12 \mG_4 \wedge  C_2 + \fr12 \mG_6 \wedge C_0  \\
        &+ \fr12 \mG_2 \wedge \mG_4-\fr12 \mG_6 \wedge \mc{G}_0+ \fr1{3!} |\t|^{-2} C_0 \mG_2 \wedge \mG_2 \wedge \mG_2,
    \end{aligned}
\end{equation}
where $\mG_p$ and $C_p$ do not transform covariantly under T-duality. However, as we show below this form of the action appears to be convenient for writing as a covariant DFT monopole action. Important here is that in comparison to \eqref{eq:WZgauge0} one finds three more terms, each gauge invariant. Note that the last term in the first line and the second term in the second line cancel each other. These has been written separately to single out the first line in the form, conventional in the literature (see e.g. \cite{Bergshoeff:2011zk}). To our knowledge, gauge invariant Wess-Zumino term for the NS5-brane has not been presented in this form in the literature. Indeed, the first line can be written by using the arguments based on gauge invariance and T-duality covariance. One the other hand each term in the second line is gauge invariant on its own and one is unable to see these using the same arguments. First line and second line transform under T-duality separately.  Although following from pretty straightforward calculation, this result seems to be crucial for writing the full effective action of NS 5-branes in a T-duality covariant form. Before turning to covariant notations, let us proceed with setting the stage for the world-volume fields of the KK5A-monopole using covariant T-duality transformation rules.

\subsection{Effective action for the KK5A-monopole}

Given isometry along a Killing vector $k^\m$ a background can be dualised using the following covariant T-duality rules
\begin{equation}
    \label{eq:covT}
    \begin{aligned}
        g'_{\m\n} & = g_{\m\n} - \fr{(\I_kg)_\m(\I_kg)_\n - K_\m K_\n}{k^2},\\
        B'_{\m\n} & = B_{\m\n} - \fr{(\I_kg)_\m K_\n - (\I_kg)_\n K_\m}{k^2},\\
        e^{2\f'} & = \fr1{k^2} e^{2\f}, \\
        C_n' & = (-1)^n \I_kC_{n+1} - C_{n-1} \wedge K - \fr{(-1)^{n-2}}{k^2}\I_k C_{n-1}\wedge K \wedge \I_kg.
    \end{aligned}
\end{equation}
Here the following notations have been used
\begin{equation}
    \begin{aligned}
        k^2& = g_{\m\n} k^\m k^\n, \\
        (\I_k \w_{p})_{\m_1\dots \m_{p-1}} & = k^{\m}\w_{\m \m_1\dots \m_{p-1}},\\
        \I_kg & = (\I_k g)_\m dx^\m = k^\m g_{\m\n} dx^\n,\\
        K & = \I_k B - d\w_0,
    \end{aligned}
\end{equation}
where $\w_0$  satisfies $\I_k d\w_0=1$  and has basically the meaning of the dualised coordinate. In the adapted frame the above reproduce the standard Buscher rules. Indeed, choosing $k^\m = \d^\m_z$ the condition $\I_k d\w_0=1$ gives $\w_0=z$. Labeling the remaining nine coordinates by small Latin indices $\m=(m,z)$ we have
\begin{equation}
    \begin{aligned}
        k^2& = g_{zz} , \\
        (\I_k \w_{p})_{\m_1\dots \m_{p-1}} & = \w_{z \m_1\dots \m_{p-1}},\\
        \I_kg & =  g_{zz} dz + g_{zm}dx^m,\\
        K & = B_{z m}dx^m - dz.
    \end{aligned}
\end{equation}
The transformations \eqref{eq:covT} then become precisely the standard Buscher rules.

Under the covariant Buscher rules the axio-dilaton and the field strength $\mG_{ij}$ transform as
\begin{equation}
    \begin{aligned}
        |\tau|&\xrightarrow{T} e^{-\phi}k^{-1}\sqrt{1+e^{2\phi}k^{-2}(C_0)^2},\\
        \mathcal{G}_{2} &\xrightarrow{T} K_{2} =  2d\omega_1 -\I_kC_3+K_1\wedge (C_1 + k^{-2} \I_kC_1\, \I_k g).
    \end{aligned}
\end{equation}
Applying these to the effective action of the NS5B-brane we obtain the following expressions for the effective action for the KK5A-monopole
\begin{equation}
    S_{DBI}^{KK5} = T_{KK5} \int d^6\sigma e^{-2\phi}k^{-2}\sqrt{1+e^{2\phi}k^{-2}(C_0)^2} \sqrt{\left|\det(G_{ij}DX^iDX^j - k^{-2}K_iK_j + \frac{e^{-2\phi}k^{-2}}{1+e^{2\phi}k^{-2}(C_0)^2}K_{ij}\right|},
\end{equation}
whdere the long derivative is define as usual as
\begin{equation}
    Dx^i = \partial x^i + k^{-2} (\partial x^j k_j)k^i.
\end{equation}
The same calculation for the Wess-Zumino term gives 
\begin{equation}
    \begin{aligned}
        S_{WZ}^{KK5} & = \int \I_k B_{7,1} - (\I_k C_5 - C_3 \wedge K - k^{-2}\I_k C_3 \wedge K \wedge \I_kg)\wedge (dc_1 - \I_k C_3 + C_1 \wedge K + k^{-2}\I_k C_1 \wedge K \wedge \I_k g )\\
        &-\fr12 (B- k^{-2}\I_k g \wedge K)\wedge dc_1 \wedge dc_1 + \fr1{3!}|\t'|^{-2} \I_k C_1 (dc_1 - \I_k C_3 + C_1 \wedge K + k^{-2}\I_k C_1 \wedge K \wedge \I_k g)^{\wedge 3}
     \end{aligned}
\end{equation}

\section{Covariant action}
\label{sec:cov}

\subsection{Full DFT monopole action}

In \cite{Blair:2017hhy} it has been shown that representatives of the T-duality orbit of the NS5-brane can be understood as projections to the physical space of a single object, called DFT-monopole, differently oriented in the dual space. At the level of background fields this has been demonstrated in \cite{Berman:2014jsa,Bakhmatov:2016kfn}, where a single solution to DFT equations of motion was shown to reproduce backgrounds of the NS5-brane, KK5-monopole and of exotic 5-branes upon a choice of physical and dual coordinates. In this section we briefly review the construction of the T-duality covariant effective action for the DFT-monopole following \cite{Blair:2017hhy} and present a covariant Wess-Zumino action, that reproduces the expressions obtained above.

We start with the section constraint of double field theory that requires the background fields to depend only on a half of coordinates, but does not specify whether these are geometric or dual. At the level of effective actions this translates into the requirement to have ten generalized Killing vectors $k_a{}^M$, where the index $a=1,\dots,10$. These are defined such as to satisfy an algebraic version of the section costraint:
\begin{equation}
    \label{eq:algsec}
    k_a{}^M k_b{}^N\h_{MN}=0.
\end{equation}
This condition simply tells that the space orthogonal to the set of ten Killing vectors at a point is an isotropic subspace of the generalized tangent space. The same construction has been used for the T-duality covariant description of D-branes in \cite{Bergshoeff:2019sfy}. The magnetic potential $D_{MNKL}$ and R-R fields couple to covariant brane charges $T_{M_1\dots M_{10}}$ and $|\l_b\rangle$ respectively. These are defined in terms of the generalized isometries as follows 
\begin{equation}
    \begin{aligned}
    T^{M_1\dots M_{10}} & = \e^{a_1\dots a_{10}}k_{a_1}^{M_1}\dots k_{a_{10}}^{M_{10}}, \\
    k_a{}^M \G_M |\l_b\rangle&=0, \\
    \fr{1}{2^5} \langle \l_b | \G^{M_1\dots M_{10}}|\l_b\rangle & = T^{M_1\dots M_{10}}.
    \end{aligned}
\end{equation}
The second line defines the brane charge $|\l_b\rangle$, while the third line fixes its normalization.  The universal expression for the brane charge is then
\begin{equation}
    |\l_b\rangle =  |\l_b| \,T_{M_1\dots M_{10}}\G^{M_1\dots M_{10}}|0\rangle,
\end{equation}
where we define $|\l_b|^2 = \langle \l_b|\l_b\rangle$. At this point it is tempting to hide  $|\l_b|$ into a normalized brane charge, which however appears to be not completely convenient. Indeed, as it will be shown below NS-NS and R-R contributions to the effective action will have the same prefactor $|\l_b|^2$ which can be hidden into the covariant tension at the last step. Using brane charge normalized to unitey will result in having different expressions for the tension in DBI and WZ actions, that might be confusing. Finally, we define a matrix $h_{ab} = k_a{}^M k_b{}^N \mH_{MN}$ and hatted derivatives of twenty worldvolume scalars along worldvolume cordinates $\s^i$
\begin{equation}
    \hdt_i Y^M = \dt_i Y^M - h^{ab}k_a{}^M k_b{}^N \mH_{NP}\dt_i Y^P,
\end{equation}
where $h^{ab}$ is the inverse of $h_{ab}$.

To construct a covariant Wess-Zumino term one needs to define interactions with R-R fields, which are combined in a DFT spinor as usual as
\begin{equation}
    |\mC\ra = \sum_p \fr{1}{p!\sqrt{2}^p} \G^{\m_1} \dots \G^{\m_p}\mC_{\m_1\dots \m_p}|0\rangle, 
\end{equation}
where the Clifford vacuum is defined as $\G_\m |0\rangle =0$. Following \cite{Blair:2017hhy} we define worldvolume forms taking values in O$(10,10)$ spinors
\begin{equation}
    \begin{aligned}
        |\cC_p\ra&= \fr{(-1)^{\fr{p(p-1)}{2}}}{p!(\sqrt{2})^p}\G_{M_1}\dots \G_{M_p} \hat{d}Y^{M_1}\wedge \dots \wedge \hat{d}Y^{M_p}  |\mC\ra,\\
        |\mG_p \ra& = d |{c}_p \ra - \fr12 \h_{MN}d\hat{d}Y^M \wedge \hat{d}Y^N \wedge |{c}_{p-3} \ra + |\mC_p\ra.
    \end{aligned}
\end{equation}
The worldvolume potentials $c_p$ are then defined as $\langle \l_b|{c}_p\rangle = |\l_b|\,  c_p$. It is important to note here, that such defined potentials $c_p$ do not transform covariantly under T-duality, i.e. are not components of an O(10,10) spinor. In contrast, $c_p$ is a $p$-form taking values in O(10,10) spinors.

In these terms the full action describing $5_2^b$-branes with $b=0,\dots,4$ then takes the form
\begin{equation}
    \label{eq:S_full_conc}
    \begin{aligned}
        S&=-\mc{T}_5 \int d^6\s e^{-2d}\hat{\tau}\sqrt{\det h} \sqrt{-\det\Big|\mH_{MN} \hdt_i Y^M \hdt_j Y^N  - \hat{\tau}^{-1}{e^d (\det h)^{-\fr14} \langle \l_b|\mG_{ij}\rangle} \Big|} \\
        &-\mc{T}_5 \int D^{M_1\dots M_4}T_{M_1\dots M_{10}}\hdt Y^{M_5}\wedge \dots \wedge \hdt Y^{M_{10}}\\
        &-\fr12\mc{T}_5\int \langle \l_b |\mG_6\rangle \wedge \langle \l_b |\mC_0\rangle-
        \langle \l_b |\mG_4\rangle \wedge \langle \l_b |\mC_2\rangle+
          \langle \l_b |\mG_2\rangle \wedge \langle \l_b |\mC_4\rangle
          -  \langle \l_b |\mG_0\rangle \wedge \langle \l_b |\mC_6\rangle\\
        & + \fr{1}{3!}\mc{T}_5 \int \hat{\tau}^{-2} \langle \l_b| \mG_2\rangle \wedge \langle \l_b| \mG_2\rangle \wedge  \langle \l_b| \mG_2\rangle    \langle \l_b| \mC_0 \rangle \\
        &+\fr12\mc{T}_5\int \langle \l_b |\mG_6\rangle \wedge \langle \l_b |\mG_0\rangle-
        \langle \l_b |\mG_4\rangle \wedge \langle \l_b |\mG_2\rangle,\\
        \hat\tau &= \sqrt{1+e^{2d}(\det h)^{-\fr12}\langle \l_b|\mC\rangle^2},
    \end{aligned}
\end{equation}
where $\mc{T}_5 = |\l_b|^{-2}T_5$ can be understood as the covariant tension. This expression completes the action of \cite{Blair:2017hhy}, which has only the first three lines above, to the full T-duality covariant action of the DFT-monopole. As it has been shown in \cite{Blair:2017hhy} the above action coupled to the full DFT action for background NS-NS fields reproduces the solution of \cite{Bakhmatov:2016kfn} to field equations. Choosing the set of generalized Killing vectors that solve the condition \eqref{eq:algsec} corresponds to choosing a T-duality frame for the solution. Equivalently, to an alternative choice of  projection to the physical space-time.

Let us now show that the above indeed reproduces the effective action for the NS5-brane in the notations of Section \ref{sec:nots} and the KK5-monopole given the covariant Buscher rules. The first is basically a repetition of the same procedure in \cite{Blair:2017hhy} but for the full action, while for the second we keep the isometry general avoiding the adapted basis. The main idea is that one is free to chose the direction of each of ten generalized Killing vectors along the dual or the normal space, assuming we have defined the corresponding split. Depending on the orientation, one ends up with different actions, matching those of the representatives of the orbit $5_2^0 -\dots - 5_2^4$. Since additionally one has to gauge fix six world-volume coordinates, actually only directions of four of ten Killing vectors matter, changing of the other will just redefine the metric.

\subsection{NS5B-brane}

Let us start with the case when all generalized Killing vectors are along the dual directions $k_a{}^M = (0, \tk_{a\m})$. For that we calculate
\begin{equation}
    \begin{aligned}
    h_{ab} &= \tk_{a\m}\tk_{b\n}G^{\m\n},\\
    T_{\m_1\dots \m_{10}} & = \det|\tk_{a\m}| \e_{\m_1\dots \m_{10}}, \\
    |\l_b\rangle& = \fr{\sqrt{\det \tk}}{2^5}\G^1\dots \G^{10} |0\rangle.
    \end{aligned}
\end{equation}
Here we understand $\tk_{a\m}$ as a 10$\times 10$ non-degenerate matrix, since there are 10 linearly independent generalized Killing vectors, which in particular implies $\det h = (\det|\tk|)^2 \det G^{-1}$ and $|\l_b|^2 = \det \tk$.

For the \textbf{DBI action} we first calculate $\hdt_i Y^M$ whose normal and dual components read
\begin{equation}
    \begin{aligned}
        \hdt_i Y^\mu & = \dt_i Y^\mu,\\
        \hdt_i Y_\mu & = B_{\mu \nu}\dt_i Y^\nu,
    \end{aligned}
\end{equation}
that gives $    \mH_{MN} \hdt_i Y^M \hdt_j Y^N = G_{\m\n}\dt_i Y^\m \dt_j Y^\n$
as expected. Now, for the R-R contribution to the DBI action we calculate
\begin{equation}
    \begin{aligned}
        \langle \l_b| \mC\rangle &= \sqrt{\det \tk} \,\mC_0,\\
        |\tmG_2 \ra& = d |{c}_1\ra - \fr{1}{4}\G_{MN} | \mC\ra \hat{d}Y^M \wedge \hat{d}Y^N.
    \end{aligned}
\end{equation}
From the second line and the definition the brane charge one easily obtains
\begin{equation}
    \begin{aligned}
        \langle \l_b|\mG_2 \ra & = \sqrt{\det 
        \tk}\Big( d\big[(\det \tk)^{-\fr12}\langle\l_b|{c}_1\rangle\big] +  \mC_{\m\n}dY^\m \wedge  dY^\n \Big) \\
        & = \sqrt{\det 
        \tk}( dc_1 + C_2 ).
    \end{aligned}
\end{equation}
Note that $c_1$ depends only on the components implicitly hidden in $|{c}_1\rangle$ and the brane charge, and does not depend on components of the Killing vector, as it should be. Finally, taking into account the relation $e^{2d}(\det h)^{-1/2} = e^{2\phi}(\det \tk)^{-1}$, we obtain for the DBI action
\begin{equation}
    S_{DBI} =  -\mc{T}_5 \int d^6 \s(\det\tk) e^{-2\phi}\sqrt{1+e^{2\phi}{ C_0^2}}\sqrt{-\det \bigg| G_{\m\n} \dt_i Y^\m \dt_j Y^\n  - \fr{e^\phi \mG_{ij}}{\sqrt{1+e^{2\phi}{ C_0^2}}}\bigg|}
\end{equation}

\textbf{Wess-Zumino} terms contain contributions from the NS-NS fields encoded in the potential $D^{MNPQ}$ and from the R-R fields encoded in the $p$-forms $|\mC_p\ra$ taking values in O(10,10) spinors . Consider first the latter and  calculate
\begin{equation}
    \begin{aligned}
        \langle \l_b|{\mC}_p\rangle & = \sqrt{\det \tk} \fr{1}{p!} C_{\m_1\dots \m_p}dY^{\m_1} \wedge \dots \wedge dY^{\m_p} = \sqrt{\det \tk} \, C_p,\\
        \fr12 \h_{MN} \hat{d} Y^M\wedge \hat{d} Y^N & = - \fr{1}{3!} H_{\m\n\r}dY^\m \wedge dY^\n \wedge dY^\r,
    \end{aligned}
\end{equation}
where form components $C_{\m_1\dots \m_p}$ in the first line are not covariant under T-diality. Hence, we have
\begin{equation}
    \begin{aligned}
         \langle \l_b| \mG_p\rangle  & = \langle \l_b| d{c}_p\rangle - \fr12 \h_{MN} d\hat{d} Y^M\wedge \hat{d} Y^N \wedge \langle \l_b| {c}_{p-3}\rangle + \langle \l_b|{\mC}_p\rangle \\
         &=\sqrt{\det \tk} \big(d {c}_p + H_3 \wedge {c}_{p-3} + C_p  \big)  = \sqrt{\det \tk}\, \mG_p,
    \end{aligned}
\end{equation}
that gives  the corresponding contributions to the Wess-Zumino term for the NS5-brane. Note, that the fields $c_p$ are not covariant under T-duality transformations, which does not contradict to the fact, that $|c_p\ra$ is a DFT spinor. Indeed, the non-covariance here is about transformation of form components $\langle \l_b|c_p\ra$, whose precise form depends on the choice of the brane charge. Finally, for the additional terms quartic in R-R fields we have
\begin{equation}
    \fr{e^{2d}(\det h)^{\fr12}}{1+ e^{2d} (\det h)^{-\fr12}\langle \l_b|\mC_0\rangle^2} \langle \l_b| \mG_2\rangle \wedge \langle \l_b| \mG_2\rangle \wedge  \langle \l_b| \mG_2\rangle    \langle \l_b| \mC_0 \rangle = \det \tk\, |\t|^{-2} {C}_0 \, \mG_2 \wedge \mG_2 \wedge \mG_2 . 
\end{equation}
Note, that the overall power of $\det \tk$ in DBI and WZ terms can be factored out and hidden in a definition of tension. Hence, we conclude that when all ten generalized Killing vectors are along dual directions the full action \eqref{eq:S_full} reproduces the standard NS5-brane action together with terms quartic in R-R forms.

\subsection{KK5A-monopole}

Consider now effective action for the KK5-monopole, that corresponds to an alternative choice of the solution to the constraint on Killing vectors. Explicitly, we have 9 generalized Killing vectors in dual directions $\tk_{\a \mu}$, $\a=1,\dots,9$, and a single Killing vector in normal directions $k^\m$. This gives the following charge
\begin{equation}
    T_{\m_1\dots \m_9}{}^\m = \e^{\a_1\dots \a_9}\tk_{\a_1\m_1}\dots k_{\a_9 \m_9} k^\m |0\rangle = \det||\tK||\e_{\m_1\dots \m_9\n}k^\m k^\n,
\end{equation}
where we defined a matrix $\tilde{K}_{a \mu}$ with components given by $\tK_{\a \m} = \tk_{\a\mu}$ and $\tK_{9 \mu} = \dt_\m \w$ with $\w$ defined by $k^\m \dt_\m \w=1$. Note, that we assume all $k's$ to be eventually constant. We used the following identity
\begin{equation}
    \label{eq:epskk}
    \e^{\a_1\dots \a_9}\tk_{\a_1\m_1}\dots k_{\a_9 \m_9} = \det||\tK||\e_{\m_1\dots \m_9\n}k^\n,
\end{equation}
which simply follows from the following observations. First, 9 dual Killing vectors $\tk_{\a}$ can be understood as contravariant components of vectors in the normal space and the remaining vector $k^\m$ orthogonal to this system completes it to a full basis in 10-dimensional (tangent) space-time. Now the LHS of \eqref{eq:epskk} is simply a 10-dimensional vector product of 9 vectors, that apparently gives a vector orthogonal to the hyperplane spanned by $\tk_\a$. This is proportional to $k^\m$. The coefficient is restored by contraction with $\e^{\m_1\dots \m_{10}}$. Given that, the brane charge reads
\begin{equation}
    \begin{aligned}
        |\l_b\rangle &= \fr{1}{2^{\fr92}9!} \sqrt{\det||\tK||}\e_{\m_1\dots \m_9 \m}k^\m \G^{\m_1\dots \m_9}|0\rangle,\\
        \langle \l_b| & = \fr{1}{\sqrt{2}} \sqrt{\det||\tK||} \langle 0| k^\m \G_\m.
    \end{aligned}
\end{equation}
Components of the matrix $h_{ab}$ then take the following form
\begin{equation}
    \begin{aligned}
        h_{\a\b} & = \tk_{\a\m}\tk_{\b\n}g^{\m\n},\\
        h_{\star \a} & = \tk_{\a \m}k^\n B_\n{}^\m,\\
        h_{\star \star } & = k^\m k^\n (g_{\m\n} - B_{\m \r}{}g^{\r\s} B_{\s\n}).
    \end{aligned}
\end{equation}
As one would expect the RHS of the above expressions can be rewritten in a more convenient form using T-dualised background metric $g'_{\m\n}$ and its inverse $g'{}^{\m\n}$. These are related by covariant T-duality transformations as in \eqref{eq:covT} and one finds
\begin{equation}
    \begin{aligned}
        g'{}^{\m\n} & = g^{\m\n}+(k^2 +K^2)k^\m k^\n + 2 k^{(\m}K^{\n)},\\
    K& = \I_{k}B - d\w.
    \end{aligned}
\end{equation}
Indices on the RHS are raised and lowered by the initial metric $g_{\m\n}$. Simple  calculation then shows
\begin{equation}
    h_{ab} = \tK_{a\m}\tK_{b\n}g'{}^{\m\n}.
\end{equation}
In particular we obtain the following components of the inverse matrices $h^{ab}$ and $(\tK^{-1})^{a\m}$ which will be useful in further calculations
\begin{equation}
    \begin{aligned}
        h^{\star \star } & = \fr{1}{k^2}, && (\tK^{-1})^{\star \m} = k^\m,
    \end{aligned}
\end{equation}
Additionally one has the following identities
\begin{equation}
    \begin{aligned}
        h^{\star \a}\tk_{\a \m} = \fr{1}{k^2}B_{\m\n}k^\n, && \det g'_{\m\n} = \det g_{\m\n} k^{-4}.
    \end{aligned}
\end{equation}

Let us start with the kinetic part of the full covariant action and show that it reproduces precisely the \textbf{DBI action} of KK5A-monopole. For that we calculate
\begin{equation}
    \begin{aligned}
        \hat{\dt}_i Y^\m & = \left(\delta_\n{}^\m - \fr{k_{\n}k^\m}{k^2}\right)\dt_iY^\n, \\
        \hat{\dt}_i \tY_\mu & = \fr{1}{k^2} k_{\m}k^\n \dt_i \tY_\n + \left(B_{\m\n} +\fr{2}{k^2}k_{[\m}B_{\n]\r}k^\r\right)\dt_i Y^\n,
    \end{aligned}
\end{equation}
where indices are raised and lowered by the metric $g_{\m\n}$ and its inverse. One notes that the expression in parentheses in the first line is simply a projector on a hyperplane orthogonal to the vector $k^\m$. This implies, that components of $Y^\n$ along the isometry vector do not appear in the action, which is the brane action manifestation of the isometric directions of the corresponding solution. On the other hand, precisely  only the component of $\tY_\m$ projected on $k^\m$ remains, while the others are simply proportional to $Y^\m$. At  the level of DFT solutions this results in dependence of the background fields on one dual coordinate. Most transparent this is seen in the adapted basis, where the only non-vanishing component of the isometry vector is $k^z=1$. Splitting the space-time index as $\m=(m,z)$ we write
\begin{equation}
    \begin{aligned}
        \hat{\dt}_iY^m & = \dt_i Y^m, && & \hat{\dt}_iY^z  &= 0, \\
        \hat{\dt}_i\tY_m & = B_{mn} \dt_i Y^n, && & \hat{\dt}_i\tY_z  &= \dt_i \tY_z.
    \end{aligned}
\end{equation}
Hence, eventually the action will contain only fields $Y^m$ and $\tY_z$, generating dependence on 3 normal coordinates and one dual coordinate after gauge fixing. This corresponds to the localized KK monopole background. We will keep the covariant notations working in a general frame and define
\begin{equation}
    \begin{aligned}
        k^\m \dt_i \tY_{\m} & = \dt_i \w,\\
        K_i & = k^\m B_{\m\n}\dt_i Y^\n - \dt_i \w,\\
        D_i Y^\m & = \dt_i Y^\m - k^{-2}k^\m k_{\n}\dt_iY^\m,
    \end{aligned}
\end{equation}
that gives for the NS-NS part of the DBI action the familiar expression:
\begin{equation}
    \mH_{MN}\hdt_i Y^{M} \hdt_j Y^N = g_{\m\n}D_i Y^\m D_j Y^\n + k^{-2}K_i K_j.
\end{equation}
Let us now turn to the contributions from the R-R fields to the DBI part of the action. For that we calculate a general expression
\begin{equation}
    \begin{aligned}
    (\det ||\tK||)^{-\fr12} \langle \l_b|\mC_{n}\rangle = &\   \I_k C_{n+1} - C_{n-1}\wedge K - \fr{1}{k^2} \I_k C_{n-1}\wedge K \wedge \I_k g,
    \end{aligned}
\end{equation}
which gives precisely the covariant Buscher rules for R-R fields. 

Hence, one observes that the fields for the choice of the brane charge $|\l_b^{KK5}\rangle$ are related to those corresponding to $|\l_b^{NS5}\rangle$  by covariant Buscher rules. This implies that the effective action \eqref{eq:S_full} reproduces exactly the action for the KK5 monopole. The same is true for the other representatives of the orbits, i.e. the $5_2^2,5_2^3, 5_2^4$ branes. In this case we have the effective actions of \cite{Kimura:2014upa}, obtained by T-dualization of the NS5-brane action.

\section{Discussion}

\label{sec:disc}

\subsection{Location, charges and invariance}

The main results we provide in the paper can be summarized by two equations. The first is the full action for the NS5B--brane in democratic  formulation
\begin{equation}
    \label{eq:fullNS50}
    \begin{aligned}
        S^{NS5B}= & -T_{5}\int d^6\xi e^{-2\phi}\sqrt{1+e^{2\phi}(C^{(0)})^2}\sqrt{-\det\left( G_{ij} - \fr{e^{\phi}}{\sqrt{1+e^{2\phi}(C^{(0)})^2}} \mathcal{G}_{ij} \right)}\\
        &- \fr12 T_5 \int_{M_6}D_6-  \mG_0 \wedge C_6  +\mG_2 \wedge C_4 - \mG_4 \wedge  C_2 + \mG_6 \wedge C_0  \\
        &\quad + \mG_2 \wedge \mG_4-\mG_6 \wedge \mc{G}_0+ \fr1{3!} |\t|^{-2} C_0 \mG_2 \wedge \mG_2 \wedge \mG_2.
    \end{aligned}
\end{equation}
The first two lines above have been presented in the literature before. The DBI action that is the first line is obtained simply by S-duality of the DBI action for the D5-brane in Type IIB theory. The first four terms in the second line are written by turning to the magnetic potential $B_6$ dual to the Kalb--Ramond field and by requiring world-volume gauge invariance. The last term in this line is gauge invariant on its own and its presence is dictated by democracy of the formalism. In this work we start with the standard formulation of the full effective action of the NS5B-brane obtained by S-duality of the D5-brane and explicitly perform the procedure of magnetic dualization of the world-volume field $c_1$ to $c_3$. This is the S-dual of the Born--Infeld vector $A$ living on the D5-brane. This allows to systematically reproduce the action above, including the last line, which avoids the analysis based on symmetries due to invariance under all relevant symmetries. To our knowledge the full Wess-Zumino term in this form has not been presented in the literature before.

The second equation is the fully O$(10,10)$-covariant expression for the action of a $5_2^b$--brane with $b=0,\dots,4$  which generalizes and completes that of \cite{Blair:2017hhy} by including the additional Wess-Zumino terms
\begin{equation}
    \label{eq:S_full}
    \begin{aligned}
        S&=-\mc{T}_5 \int d^6\s e^{-2d}\hat{\tau}\sqrt{\det h} \sqrt{-\det\Big|\mH_{MN} \hdt_i Y^M \hdt_j Y^N  - \hat{\tau}^{-1}{e^d (\det h)^{-\fr14} \langle \l_b|\mG_{ij}\rangle} \Big|} \\
        &-\mc{T}_5 \int D^{M_1\dots M_4}T_{M_1\dots M_{10}}\hdt Y^{M_5}\wedge \dots \wedge \hdt Y^{M_{10}}\\
        &-\fr12\mc{T}_5\int \langle \l_b |\mG_6\rangle \wedge \langle \l_b |\mC_0\rangle-
        \langle \l_b |\mG_4\rangle \wedge \langle \l_b |\mC_2\rangle+
          \langle \l_b |\mG_2\rangle \wedge \langle \l_b |\mC_4\rangle
          -  \langle \l_b |\mG_0\rangle \wedge \langle \l_b |\mC_6\rangle\\
        & + \fr{1}{3!}\mc{T}_5 \int \hat{\tau}^{-2} \langle \l_b| \mG_2\rangle \wedge \langle \l_b| \mG_2\rangle \wedge  \langle \l_b| \mG_2\rangle    \langle \l_b| \mC_0 \rangle \\
        &+\fr12\mc{T}_5\int \langle \l_b |\mG_6\rangle \wedge \langle \l_b |\mG_0\rangle-
        \langle \l_b |\mG_4\rangle \wedge \langle \l_b |\mG_2\rangle,\\
        \hat\tau &= \sqrt{1+e^{2d}(\det h)^{-\fr12}\langle \l_b|\mC\rangle^2},
    \end{aligned}
\end{equation}
where $\mc{T}_5 = |\l_b|^{-2}T_5$ can be understood as the covariant tension. We check that this reproduces the action \eqref{eq:fullNS50} upon a projection, defined by a choice of (generalized) Killing vectors $k_a{}^M$. The covariant charge of the object is encoded in 
\begin{equation}
    \label{eq:defs_conc}
\begin{aligned}
    T^{M_1\dots M_{10}} & = \e^{a_1\dots a_{10}}k_{a_1}^{M_1}\dots k_{a_{10}}^{M_{10}}, \\
    |\l_b\rangle &= \alpha T_{M_1\dots M_{10}}\G^{M_1\dots M_{10}}|0\rangle, \\
    k_a{}^M \G_M |\l_b\rangle&=0, \\
    k_a{}^M k_b{}^N\h_{MN}&=0,
    \end{aligned}
\end{equation}
with the prefactor $\a$ depending on generalized Killing vector components (see in the main text). 

Generalized Killing vectors are chosen such as to satisfy the algebraic condition in the last line of \eqref{eq:defs_conc}, that in principle has many solutions. This gives ten isometry directions, which are always assumed to be orthogonal to the brane. This implies that by gauge fixing one has additionally six isometric directions corresponding to world-volume of the brane. As it has been discussed in greater details in \cite{Berman:2014jsa,Bakhmatov:2016kfn,Blair:2017hhy} to compare solutions of DFT equations to space-time backgrounds one has not only to solve the section constraint, that leaves dependence on at most ten coordinates, but also identify which coordinates are geometric. In other words, we need to fix a set of ten coordinates that define interval in the physical space-time, which does not necessary has to coincide with the set of ten coordinates on which a background depends. In the context of the present paper this is reflected in the possibility to choose ten generalized Killing vectors $k_a{}^M$, which fix the set of directions which the corresponding DFT background does not depend on. Additionally, one has six isometries along physical coordinates, to end up with a 5-brane. The remaining four coordinates can be either physical or dual leaving us with five choices corresponding to branes $5_2^0,\dots,5_2^4$.

\subsection{Relation to other approaches}

Since the literature is widely populated by papers presenting effective actions for NS 5-branes in various forms, we find it useful to list differences between these results and the present paper. The main point is that the approaches present in the literature seem to cover only a part of the full picture, while the effective action developed here claims to be the full and complete expressions. Let us present the (incomplete) list of known approaches in more details.

\begin{enumerate}
    \item In the original paper \cite{Eyras:1998hn} the effective action for the NS5B-brane has been presented by T-dualisation of the KK5A-monopole action. The latter in turn has been obtained by dimensional reduction of the KK6-monopole of the 11-dimensional theory. 
    \item In the work \cite{Bergshoeff:2011zk} gauge invariant Wess-Zumino terms for solitonic 5-branes (including exotic 5-branes) in a democratic formulation have been presented. The paper only considers the gauge-dependent part of Wess-Zumino terms.
    \item Subsequent papers \cite{Chatzistavrakidis:2013jqa, Kimura:2014upa} considered full effective actions for all branes on the NS5B-orbit in a formulation that is not manifestly covariant under T-duality.
    \item In the work \cite{Blair:2017hhy} the formalism of generalized Killing vectors associated to the brane orientation has been introduced, which allowed to write dwon the  effective DBI action in a democratic form completely covariant under T-duality. Certain steps have been made towards the covariant Wess-Zumino action and the result of \cite{Bergshoeff:2011zk} has been rewritten in the suggested formalism.
    \item In the work \cite{Bergshoeff:2019sfy} a covariant formulation of the Wess-Zumino action for solitonic 5-branes and $g_s^{-\alpha}$ branes with $\alpha >2$ have been constructed in the split form. Hence, only the internal sector of the split DFT fields has been taken into account.
\end{enumerate}

Although being complete, T-duality covariant and democratic the presented effective action does not mark the end of the story. Quite the opposite, it raises several interesting questions to research. The most straightforward is whether the formalism of generalized Killing vectors developed in \cite{Blair:2017hhy} is valid for any branes of string theory, not only the solitonic ones. As an example one could try to apply the formalism to Dp-branes to construct an effective action for a single D-object, whose orientation is defined by the choice of Killing vectors. The corresponding DFT background would always depend on ten (nine, if the dual time is to be avoided) coordinates, only a subset of which is physical. Similar idea has been suggested in \cite{Bergshoeff:2019sfy} which however faced the problem of the $B_2$ field in the DBI action. Another approaches developing a T-duality covariant description of Dp-branes from different perspectives present in the literature. In the works \cite{Albertsson:2008gq,Albertsson:2011ux,Ma:2015yma, Blair:2019tww}  boundary conditions are defined for a T(U)-duality covariant open string which on the one hand allows to calculate beta-function for (constant) background fields in the covariant formalism, and one the other hand to define Dp-branes as particular subspaces in the generalized doubled (exceptional) space. Along these line in the work \cite{Asakawa:2012px} Dp-branes have been shown to correspond to Dirac structures on the subspace. Recently a similar approach based on Born sigma-models and para-Hermitian geometry has been considered in \cite{Marotta:2022tfe}, where Dp-branes are identified with maximally isotropic vector bundles. Interesting is that these might not admit the standard geometric picture in terms of submanifolds, which could be a signature of possible dependence on dual coordinates, relating this approach to the picture based on generalized Killing vectors.

The important feature of the fully covariant effective action is that it allows background fields to depend on dual coordinates for certain choices of generalized Killing vectors. E.g. this is possible for the KK5A-monopole, that is characterized by a single isotropy in the (transverse) physical space. On the one hand, as it has been shown in \cite{Bergshoeff:2019sfy} naively this dependence spoils gauge transformations of the corresponding flux making it not gauge invariant. On the other hand pure sigma-model calculation of \cite{Harvey:2005ab} without referring to the formalism of double field theory shows that dependence on the dual coordinate is indeed a physical effect related to world-volume instantons, rather than a DFT artifact. Hence, the important check that the effective action \eqref{eq:S_full_conc} indeed describes dynamics of an object in the full doubled space would be its invariance under gauge transformations with all possible dependence on doubled coordinates allowed.

\section* {Acknowledgments}

The work of EtM has been supported by the Foundation for the Advancement of Theoretical Physics and Mathematics ``BASIS'', grant No 21-1-2-3-1. The authors thank Fabio Riccioni for numerous useful discussions and for providing detailed feedback on drafts of this work.

\appendix

\bibliography{bib.bib}
\bibliographystyle{utphys.bst}

\end{document}